# Simultaneous Spectroscopic and Photometric Observations of Binary Asteroids

D. Polishook[1,2], N. Brosch[2], D. Prialnik[1] and S. Kaspi[2]

[1]Department of Geophysics and Planetary Sciences, Tel-Aviv University, Tel-Aviv 69978, Israel

[2]The Wise Observatory and the Raymond and Beverly Sackler School of Physics and Astronomy, Tel-Aviv University, Tel-Aviv 69978, Israel.

[*] Corresponding author: David Polishook, david@wise.tau.ac.il

**Abstract**

We present results of visible wavelengths spectroscopic measurements (0.45 to 0.72 microns) of two binary asteroids, obtained with the 1-m telescope at the Wise Observatory on January 2008. The asteroids *(90) Antiope* and *(1509) Esclangona* were observed to search for spectroscopic variations correlated with their rotation while presenting different regions of their surface to the viewer. Simultaneous photometric observations were performed with the Wise Observatory's 0.46-m telescope, to investigate the rotational phase behavior and possible eclipse events.

*(90) Antiope* displayed an eclipse event during our observations. We could not measure any slope change of the spectroscopic albedo within the error range of 3%, except for a steady decrease in the total light flux while the eclipse took place. We conclude that the surface compositions of the two components do not differ dramatically, implying a common origin and history.

*(1509) Esclangona* did not show an eclipse, but rather a unique lightcurve with three peaks and a wide and flat minimum, repeating with a period of 3.2524 hours. Careful measurements of the spectral albedo slopes reveal a color variation of 7 to 10 percent on the surface of *(1509) Esclangona*, which correlates with a specific region in the photometric lightcurve. This result suggests that the different features on the lightcurve are at least partially produced by color variations and could perhaps be explained by the existence of an exposed fresh surface on *(1509) Esclangona*.

**Introduction**

Current models suggest that some binary asteroids are created from disrupted bodies (Bottke et al. 2002; Walsh and Richardson 2006; Durda et al. 2004). Although the disruption can occur due to different causes, such as too fast rotation over the rubble pile barrier, tidal forces due to close encounters with planets (in the case of NEAs) or a catastrophic collision, all scenarios end up with the exposure of fresh surfaces. This younger area is less altered by space weathering effects, such as solar wind, cosmic rays and micrometeorite bombardment that modify the optical characteristic of the surface (Clark et al. 2002), thus these exposed fresh areas may not present the dark and redder reflectance spectra as other asteroid surfaces do. This is true if the asteroid's age as a binary is shorter than the timescale of space weathering effects which is not obvious and is estimated between $5 \cdot 10^4$ to $10^8$ years (Chapman 2004).

Finding such areas, by detecting a change in the surface's colors, can help determine the timescale of space weathering effects or pinpoint the time of disruption of the progenitor asteroid. Observing fresh areas allows a glimpse to the asteroids' interior and a possibility to compare it with exposed material on the surface. The model of "satellite accretion" by Walsh et al. (2008) predicts that a single exposed fresh area should not exist; rather, the primary should lose material from its equator due to spin-up, thus a color change will be seen only for asteroids with inclined axial orientation. Nonetheless, if there is a big difference in colors between the two components of the binary, the disruption model could be questioned and other models for binaries formation could be adopted, such as satellite capturing (Richardson and Walsh 2006). To conclude, binary asteroids are good candidates to find color variation and to link between different aspects of asteroids evolution such as disruption and space weathering effects. In this work, we've searched for such color variation on the surfaces of two binary asteroids by spectroscopic measurements.

**Work method**

The measurement of reflectance spectra is the most effective remote-sensing technique to characterize the surface composition of many asteroids and is the key method to classify asteroids (Bus and Binzel 2002). Rotationally resolved spectra are also used to determine any variance in the asteroid's surface composition while it spins around its axis (Gaffey 1997, Mothé-Diniz et al. 2000, Rivkin et al. 2006).

Recently, the young asteroid *(832) Karin*, formed ~5.75 Myr ago by a collisional breakup (Nesvorný et al. 2002), was the issue of a dispute regarding spectral variations on its surface (Sasaki et al. 2004, Chapman et al. 2007, Vernazza et al. 2007). The current consensus is that no such variations exist on *(832) Karin*.

Color variation between fresh and old surface start to appear on the slope of the continuum spectra around 0.6 μm and increases with higher wavelength (see Fig. 5 at Chapman 2004). Therefore, spectroscopic observations at the visible wavelength range might show subtle color variations even though better results can be obtain by observations at the near IR. The near IR holds more prominent differences in an asteroid's spectrum, such as the depth of the absorption band in 1 μm (Clark et al. 2002).

Searching for spectral differences between the two components of a binary asteroid, we confronted two problems: *i*) because our observing equipment (the 1-m telescope) is not equipped with adaptive optics, the binary components cannot be resolved, and *ii*) if color variations were found, we would not be able to specify the rotation phase where the color change occurs and would not distinguish between spectral variations and shape variations that appear in the lightcurves. Confronting the first problem can be done when the binary's components eclipse each other especially in the case of a total eclipse when only the reflectance spectrum of one component is visible. Total and mutual eclipses are more likely to take place if the diameter ratio of the binary is close to unity than in the case of a big asteroid with a small satellite, and allows the independent measurement of both components. Photometric observations provide an independent means of confirming the occurrence of an eclipse and yields more rotation phase information, such as brightness changes due to rotation of the elongated body. Therefore, to solve the second problem, we performed simultaneous photometric observations using a second telescope while the asteroids were spectroscopically monitored. Even though this observing technique is limited to specific and therefore rare geometric conditions, it can be used with relatively small telescopes lacking adaptive optics. Therefore, we encourage the community to perform observations aimed at rotationally resolving the spectra of binary asteroid.

**Observations, reduction, measurements and calibration**

This paper focuses on simultaneous spectroscopic and photometric observations of two asteroids, *(90) Antiope* and *(1509) Esclangona,* performed on January 11[th] to 14[th], 2008. Some previous photometric observations were collected on January 7[th] and 8[th]. The observational circumstances of these asteroids are summarized in Table I, which lists the asteroid's designation, the observation date, the time span of the observation during that night, the object's heliocentric distance (r), geocentric distance (Δ), and phase angle (α). In addition, the object's mean observed magnitude is listed, followed by the number of obtained spectra for each night.

*Spectroscopy*

The observations were performed at the Wise Observatory (code: 097, E 34:45:47, N 30:35:46). We used the observatory's 1-m Ritchey-Chrétien telescope equipped with the Faint Object Spectrographic Camera (FOSC, see Kaspi et al. 1995) and a cryogenically-cooled Princeton Instruments (PI) CCD for obtaining spectroscopic measurements with a plate scale of ~1.75 arcsec per pixel. The FOSC contains two sets of grisms: a 600 line per mm grism which gives a spectral dispersion of 3.17 Å per pixel and a 300 line per mm grism with a spectral dispersion of 6.34 Å per pixel. The effective wavelength range of both sets for the studied asteroids was 4500-7200 Å.

In order to keep the asteroids inside the slit, the spectrograph was rotated to align the slit with the asteroid track on the sky. This setting causes the asteroid to travel along the slit and no data is lost due to the asteroids' sky motion. To avoid light losses, since the slit was not at the parallactic angle, a slit of 10" width and 10.2' length was used and the asteroids were observed at air masses below 1.6. The exposure time was between 900 to 1200 seconds, and auto-guiding was used. The S/N ranged between 30 and 60, and the resulted flux scattering was between 3% to 5% of the flux value.

The images were reduced in the standard way using the *IRAF* software. Ten bias frames were taken each night. Short exposures of the twilight sky and an internal halogen lamp were used to flat-field the data. The twilight flats were used to calibrate the spatial illumination pattern along the slit, while the internal lamp flats were used to calibrate the pixel-to-pixel sensitivity variations of the CCD detector. For wavelength calibration, we also regularly obtained He-Ar spectra. The extraction width was 8-10 pixels and included all the light from the asteroid as it advanced along the slit. The spectra were flux-calibrated using spectrophotometric standard stars observed on January 13[th] and were corrected for extinction using the mean extinction curve of the Wise Observatory. Cosmic rays were removed manually. To calibrate the relative reflectance of each asteroid, spectra of the solar analog Hyades 64 (HD28099, V=8.1 mag., G2V) were also obtained every night. This widely accepted solar analog (Hardorp 1978) was observed in an air mass of 1.2 and was measured and calibrated in an identical way to the asteroids. To retain the real ratio between the reflectance spectra at different epochs, a reference spectrum was normalized to unity at 5500 Å and all the other spectra were normalized by the same value. To verify the derived spectra we compared them to the literature: *(90) Antiope* was measured by SMASS (Bus and Binzel 2002), and (1509) Esclangona by ECAS (Zellner et al. 1985). The matches are good and appear in Fig. 1.

Because our useful spectra range did not reach above 7200 Å, a match of a linear fit to each reflectance spectra was sufficient to describe the reflectance slope and the slope's error. The fit was done using a polynomial fitting with one degree (Press et al. 1989). The reflectance slopes were compared and differences were searched between them.

*Photometry*

In parallel to the spectroscopic observations, photometric measurements were performed with an 18" Centurion telescope (hereafter *C18*) located in a separate dome, 50 meters away from the 1-m telescope. This telescope is operated in a semi-automated mode and requires minimal attention from the astronomer (see Brosch et al. 2008 for the telescope's description). An SBIG ST-10XME CCD was used at the f/2.8 prime focus. This CCD covers a field of view of 40.5'x27.3' with 2184x1472 pixels, each pixel subtending 1.1 arcsec, and is used in white light. The asteroids were observed while crossing a single field per night, thus the same comparison stars were used while calibrating the images.

The images were reduced in the standard way using bias and normalized flatfield images. We used the IRAF *phot* task for the photometric measurements. The photometric values were calibrated to a differential magnitude level using local comparison stars (around ~300 stars per field) followed by absolute calibration using Landolt standards (Landolt 1992). In addition, the asteroid magnitudes were corrected for light travel time and were reduced to a 1 AU distance from the Sun and the Earth to yield reduced R values (Bowell et al. 1989). Refer to Polishook and Brosch (2009) for detailed description of the photometric procedures of observation, reduction and calibration using the *C18*.

For a consistency check, we compared the photometric and the spectroscopic measurements. The logarithm of the flux in the wavelength range 4500 to 7200 Å of each spectrum was plotted on the photometric lightcurve at the time of mid-exposure of the spectroscopic measurement, and the match between the two curves is evident in Figures 2, 6 and 9.

**Results and discussion**

*(90) Antiope*

The binary nature of *(90) Antiope*, which is located in the outer main belt (a=3.16 AU), was discovered by Merline et al. (2000) using adaptive optics. Two characteristics of this binary asteroid are unusual: *i)* its components rotate in a synchronous motion of 16.505 hours; and *ii)* both components have similar diameters (of 91 and 86 km; Descamps et al. 2007); these make *(90) Antiope* a good candidate for testing our work method, even though it is not clear if fresh surface is exposed on a binary of this size.

Descamps et al. (2007) performed an extensive campaign on *(90) Antiope* using photometry and adaptive optics and succeeded to model the orbital parameters of the components and to predict circumstances of mutual eclipses. The 171 km separation between the components enabled them to calculate a density of 1.25±0.05 g/cm$^3$ and to derive the spherical shape of the asteroid's components. The model of the components' orbits around the center of mass allowed Descamps et al. to identify appearances of mutual eclipses on the lightcurve. Fig. 9 of Descamps et al. (2007) demonstrates how a specific orbital position matches their measured lightcurve. The unique deep V-shaped minima and shallow U-shaped maxima, which are typical of eclipsing events of synchronous binaries (Polishook and Brosch 2008), are easily seen.

Our photometric measurements of *(90) Antiope* from January 11$^{th}$ yielded a lightcurve (Fig. 2) similar in shape, frequency and amplitude to that of Fig. 10 in Descamps et al. (2007). Since their lightcurve matches an almost total eclipse (see the right panel of Fig. 10 in Descamps et al.), we conclude that our observations on January 11$^{th}$ were also performed close to a full eclipse. This implies that only a single component was spectroscopically measured at this epoch.

The derived spectra is flat as expected of a C-type asteroid, which is *(90) Antiope*'s classification (Bus and Binzel 2002). All of the 12 spectra obtained on January 11$^{th}$ are presented in Fig. 3 and were shifted in the Y-axis for display. Fig. 4 displays a few examples of these spectra in a real ratio of the spectral albedo to show that they were taken before, during and after the eclipse, as marked on the lightcurve in Fig. 2. The spectral slopes do not change within the error range of the reflectance spectrum and remain -0.11±0.01 albedo per one micron, thus we conclude that the reflectance spectrum of *(90) Antiope* is constant within 3% of the flux value.

*(90) Antiope* was observed again on the following night, January 12$^{th}$, to detect the secondary eclipse after 1.5 rotations (the components complete one rotation in ~16 hours). All of the 12 spectra obtained on January 12$^{th}$ are presented in Fig. 5 and were shifted in the Y-axis for display. The mid-exposure time of the reflectance spectra are marked on the corresponding lightcurve (Fig. 6) and are all normalized by the same factor. Once again, no significant slope change is measured within 3%. Fig. 7 compare between two spectra from the two nights, showing a similarity between the two components' spectra.

The constant slope of the reflectance spectra for the two components of *(90) Antiope* suggests they have the same material on their surfaces, and that this material underwent the same physical processing. This might indicate a common origin for the two components, a conclusion that is consistent with a recent study by Descamps et al. 2009.

*(1509) Esclangona*

*(1509) Esclangona* is an S-type asteroid (Tholen and Barucci 1989), belonging to the Hungaria group (a=1.866 AU) with a spin period of 3.247 hours (Warner 2005). Its binary nature was discovered by Merline et al. (2003) using direct imaging with ESO's VLT at a projected separation between the components of about 140 km. A measurement with the IR Astronomical Satellite (IRAS) approximates the primary diameter as 8.2±0.6 km (Tedesco et al. 2004) while adaptive optics observations with the Keck II telescope estimate it as 12 km (Merline et al. 2003b). Merline et al. 2003 suggested that the companion has a diameter of about 4 km, given the difference in magnitudes.

Our simultaneous spectroscopic and photometric observations of *(1509) Esclangona* performed on January 14$^{th}$ were preceded by photometric observations on 7$^{th}$ and 8$^{th}$ of January. To retrieve the lightcurves frequency and amplitude, data analysis included folding the data points using the Fourier series to determine the variability period (Harris and Lupishko 1989). The best match chosen by least squares is 3.2524±0.0003 hours, about 20 seconds longer than measured by Warner (2005).

The folded lightcurve from the three nights (Fig. 8) of *(1509) Esclangona* presents unique features without the usual symmetric two-peak pattern: it consists of two peaks with amplitudes of 0.06 and 0.08 *mag.* followed by a wide and flat minimum (~25% of the period) and a large amplitude peak of about 0.16 *mag.* Although these features did not appear on the lightcurve published by Warner, we completely trust them as real. We observed them at three different nights with an error of less than 0.01 mag (compared to a spread of 0.05 mag on Warner's lightcurve). We also note that these features might be visible in a specific viewing geometry, like in the case of an eclipse. However, the deep minimum in the lightcurve is clearly not an eclipse event because it repeats itself every 3.2524 hours; this is a very short time compared to the ~100 years it takes a satellite with a mean separation of 140 km to complete one orbit. If an eclipse cannot explain the unique lightcurve, could changes *in (1509) Esclangona*'s colors explain it?

Fig. 9 presents the lightcurve of *(1509) Esclangona* from January 14$^{th}$, with the corresponding average values of the reflectance spectra drawn on it. The spectra are normalized by the same factor, hence they are displayed in real ratio relative to each other. As seen in Fig. 9, the average values of each spectrum fit the brightness variation of the asteroid's lightcurve, verifying the measurements and calibration processes that were done on the spectra. The reflectance spectra themselves are presented in Fig. 10, shifted on the Y-axis for display.

The slope of the reflectance spectra of (1509) Esclangona is slightly changing. An example is displayed in Fig. 11 that presents two reflectance spectra with different slopes and which demonstrate the alteration in the asteroid's color. These two reflectance spectra were superimposed in the blue side to clear the slope difference on the red side. The slopes values (scaled to fit the Y axis for display) are plotted on the lightcurve (Fig. 12) that shows a changing slope of the reflectance spectra as *(1509) Esclangona* rotates. The decrease of the slope is between 7 to 10 percent of the mean of the slopes values. To corroborate this, we note that the decrease and increase of the slope are continuous and fit a specific phase of the lightcurve – the wide and flat minimum. To reject the possibility of a slope change due to atmospheric effects we plot (Fig. 13) the slope values against the air mass of each spectrum and show that it is not correlated with the telescope elevation. To refute the possibility of thin clouds changing the reflectance spectra, we plotted the photometric lightcurves of some reference stars that were located in the same field of view of the *C18*. These lightcurves (Fig. 14) were not calibrated, hence they reflect the atmospheric stability. The brightness of the nearby stars remained constant while the slope of the reflectance spectra changed (compare with Fig. 12), supporting the lack of clouds. However, we should note that *(1509) Esclangona* was only spectroscopically measured during one rotation and more spectroscopic data would be needed to confirm that the slope change repeats itself periodically.

The correlation between a major feature on the photometric lightcurve and the alteration of the spectral slope connects between shape and color. Different configurations could explain this linkage and we point out some ideas: *i*) *(1509) Esclangona* has a peanut-like shape which made it, in the specific observed geometry, cast a shadow on itself, thus creating the unique patterns of the lightcurve. The bluer area can be explained by a colored spot that was created by a big crater or by dust scattered following an impact. However, this scenario does not consider the linkage between these two phenomena and does not relate to the binary nature of *(1509) Esclangona*. *ii*) One of the narrower sides of *(1509) Esclangona*'s primary component is flatter compared to the opposite narrow side of the asteroid which has a bullet shape. This flat side, which corresponds to the wide minimum of the lightcurve, is bluer than the rest of the surface, as the reflectance spectra measurements show. This flat and blue side was recently exposed when the asteroid was disrupted to become a binary, and therefore its surface was not yet altered by space weather effects as the other parts of the asteroid. This idea can be examined by more photometric measurements at different phase angles and apparitions that will reveal *(1509) Esclangona*'s model shape and axis orientation. *iii*) Due to the large separation between the primary and the secondary of *(1509) Esclangona* (between 23 to 35 $R_p$, Merline et al. 2003b), Durda et al. (2004) suggested that it is an Escaping Ejecta Binary (EEB), meaning both components are fragments that escaped a catastrophic collision on similar trajectories and they became bound. This suggests that *(1509) Esclangona* has a collisional history that could explain the exposure of a fresh and unweathered surface, and an asymmetric shape. Either way, more observations of *(1509) Esclangona* are therefore needed in order to determine its true evolution history.

**Acknowledgements**

David Polishook is grateful for an *Ilan Ramon* doctoral scholarship from the Israel Space Agency (ISA) and the Ministry of Science, Sports and Culture. We thank the Wise Observatory staff for their continuous support. The research was supported by the Israeli Ministry of Science, Culture and Sport that created a National Knowledge Center on NEOs and asteroids at the Tel-Aviv University.

**Tables:**

**Table I:** Observation circumstances: asteroid designation, observation date, nightly time span of the specific observation, the object's heliocentric (r), and geocentric distances (Δ), the phase angle (α), and the average magnitude (mean reduced R). The number of obtained spectra appears at the last column.

| Asteroid name | Date | Time span [hours] | r [AU] | Δ [AU] | α [Deg] | Mean R [Mag] | Number of spectra |
|---|---|---|---|---|---|---|---|
| *(90) Antiope* | Jan 11, 2008 | 9.1 | 3.6 | 2.6 | 2.5 | 13.3 | 12 |
|  | Jan 12, 2008 | 7.2 | 3.6 | 2.6 | 2.8 | 13.3 | 12 |
| *(1509) Esclangona* | Jan 7, 2008 | 4.8 | 1.8 | 1.0 | 22.4 | 14.8 | - |
|  | Jan 8, 2008 | 1.5 | 1.8 | 1.0 | 22.1 | 14.8 | - |
|  | Jan 14, 2008 | 5.3 | 1.8 | 1.0 | 19.9 | 14.6 | 13 |

**Figures:**

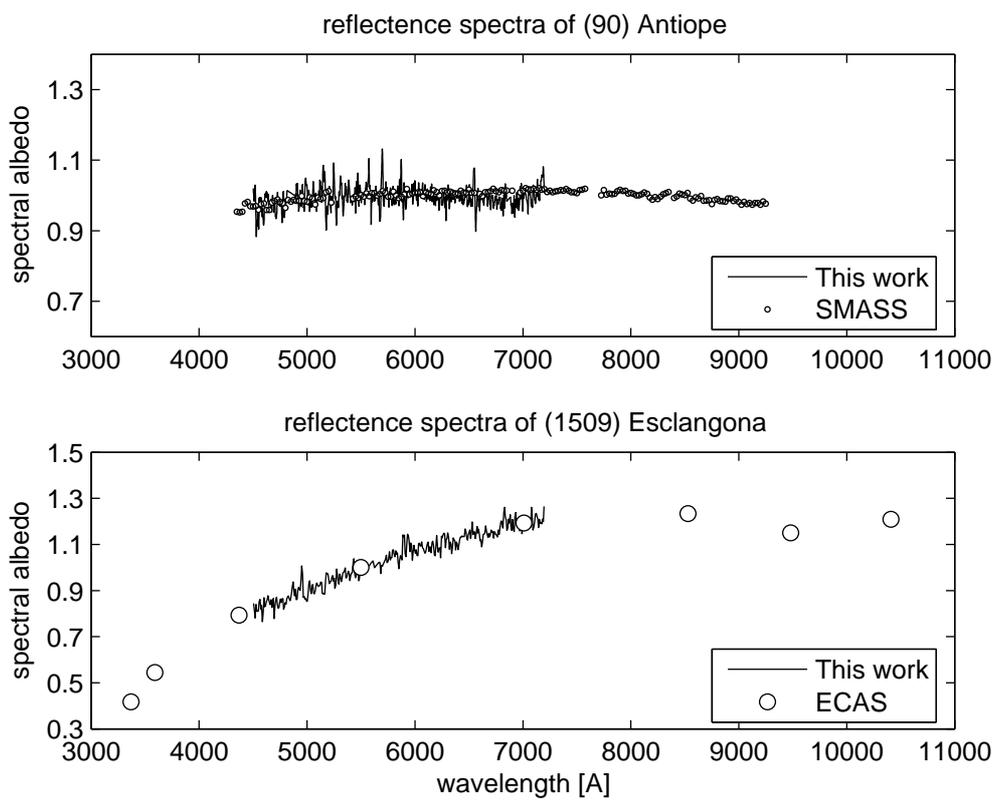

Fig. 1.

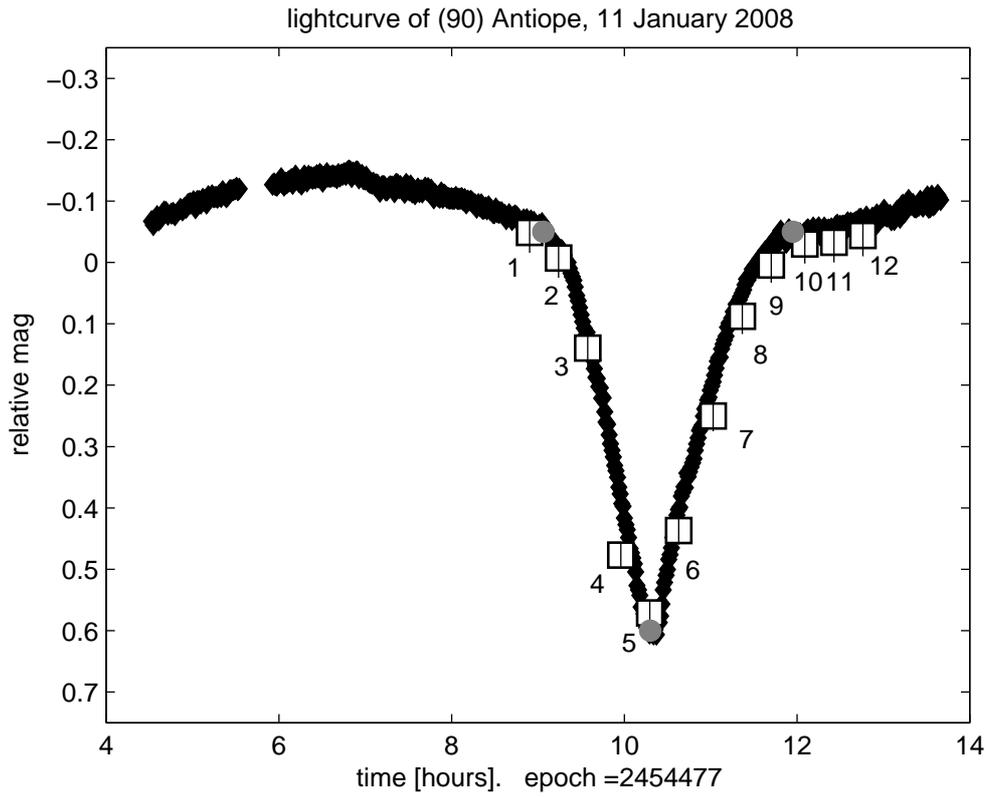

Fig. 2.

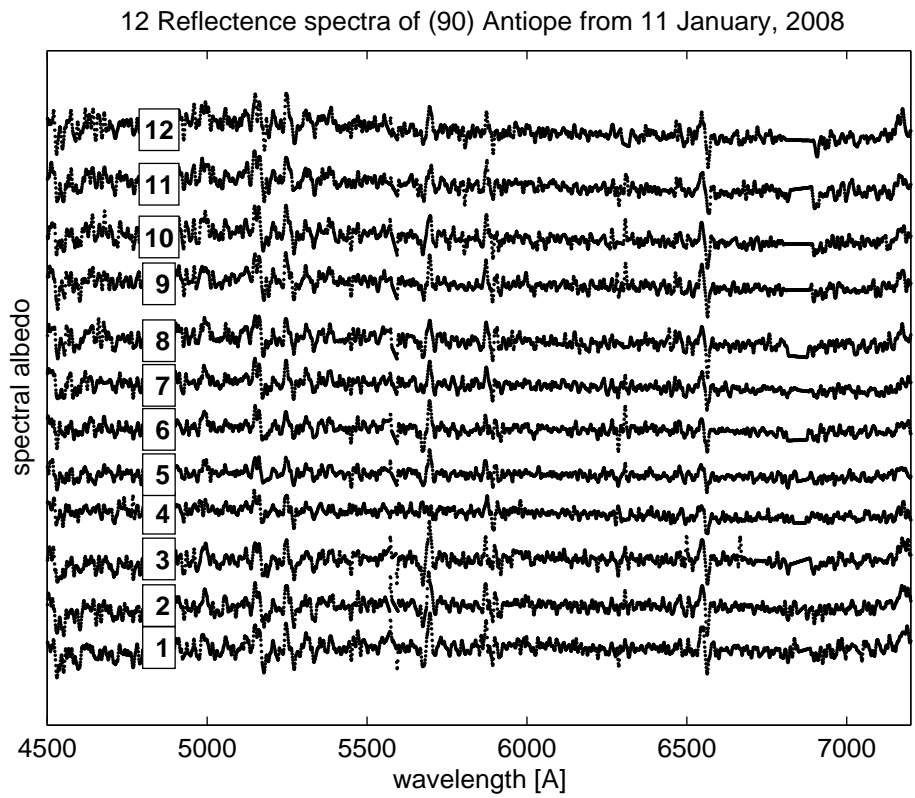

Fig. 3.

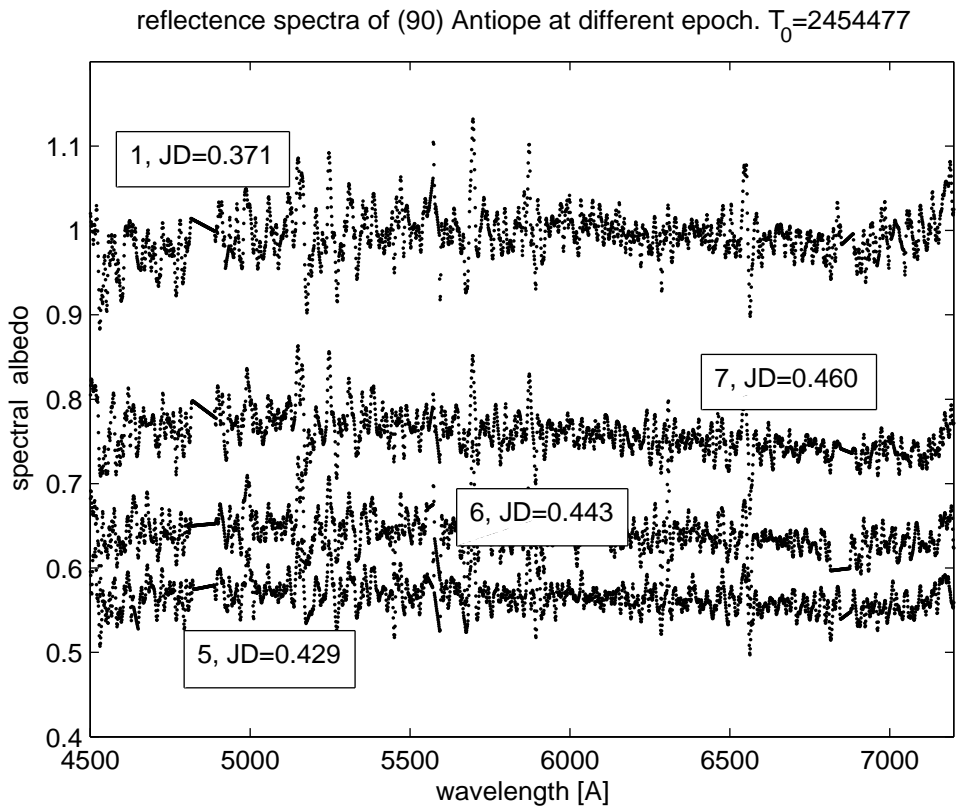

Fig. 4.

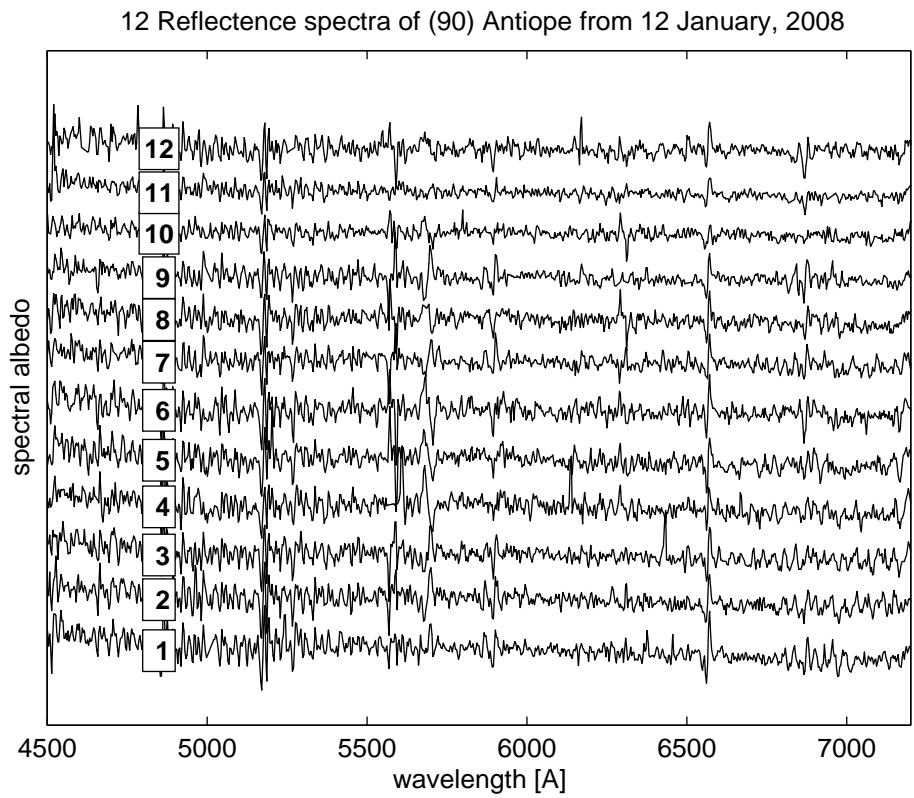

Fig. 5.

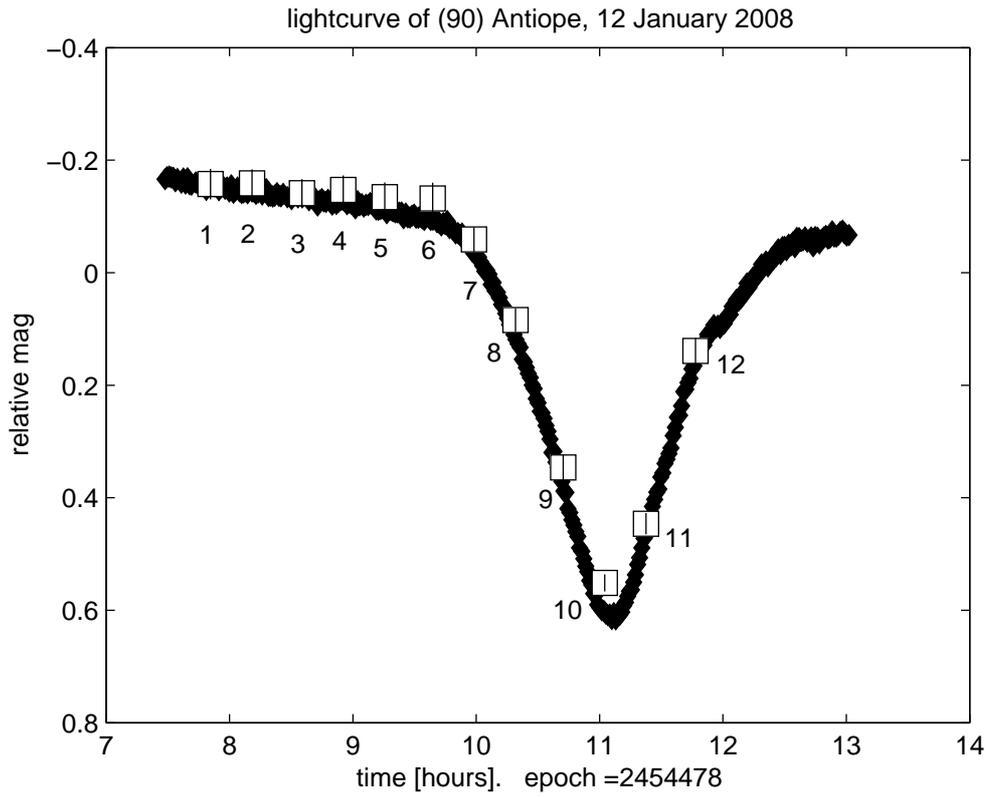

Fig. 6.

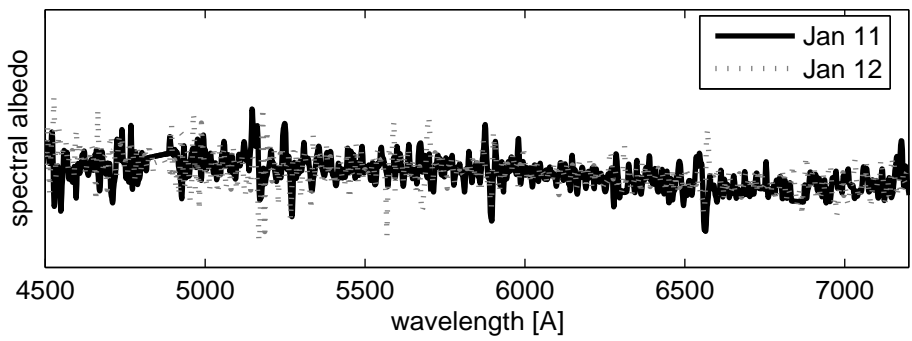

Fig. 7.

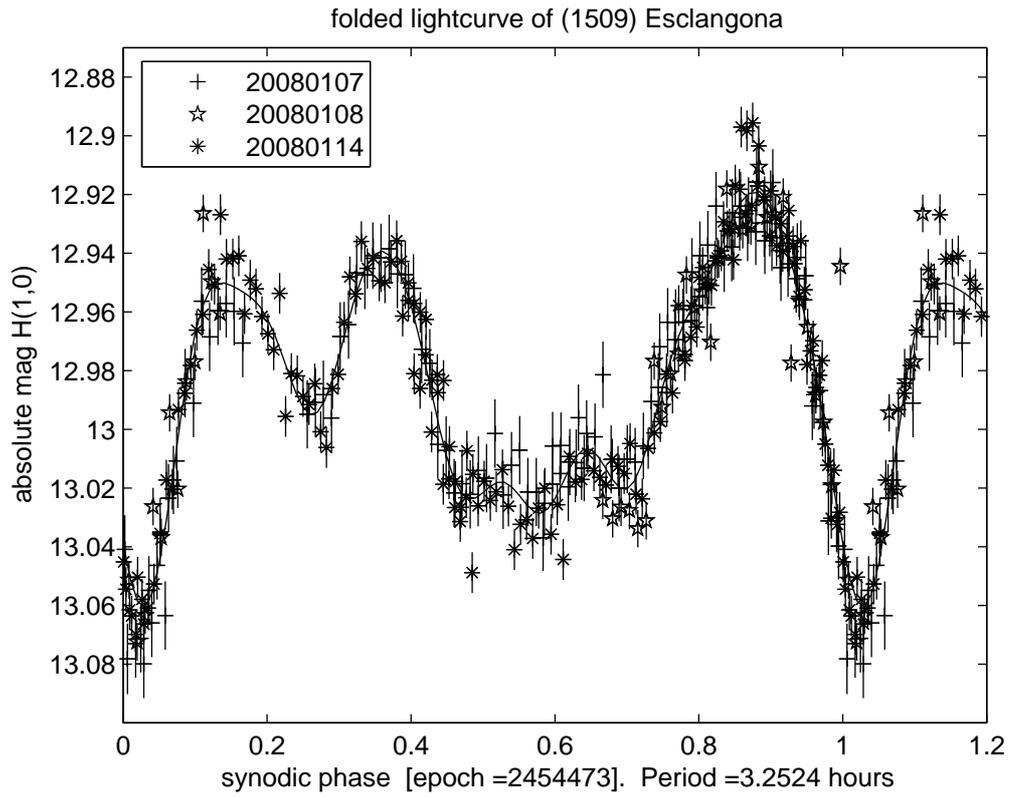

Fig. 8.

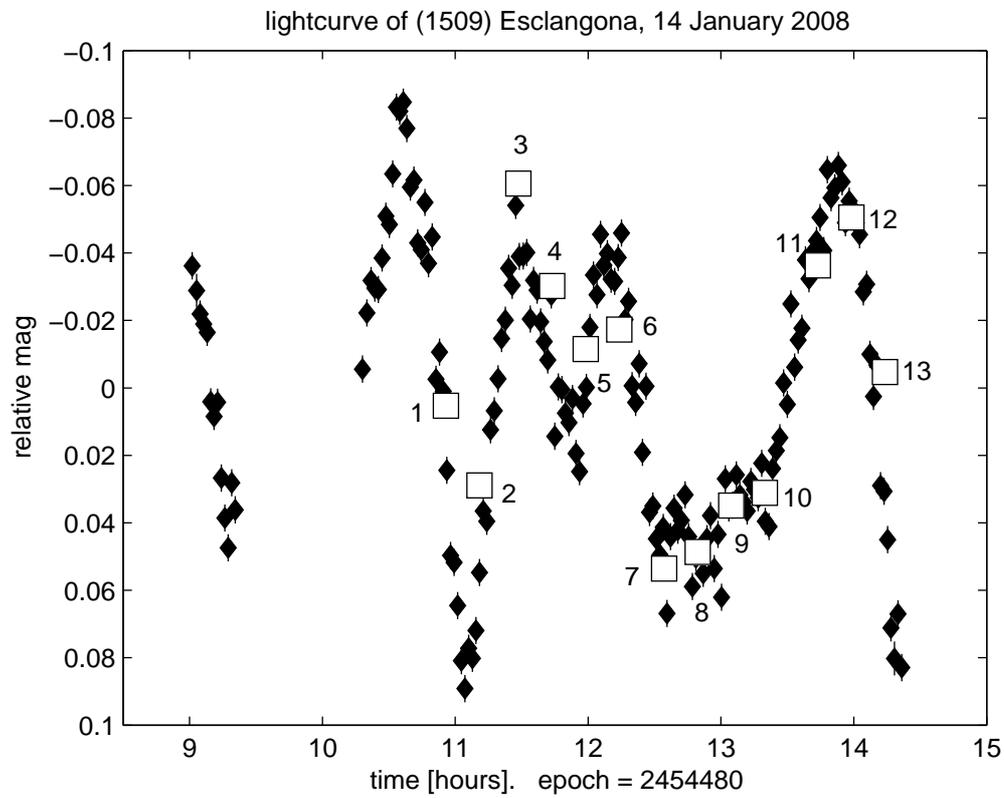

Fig. 9.

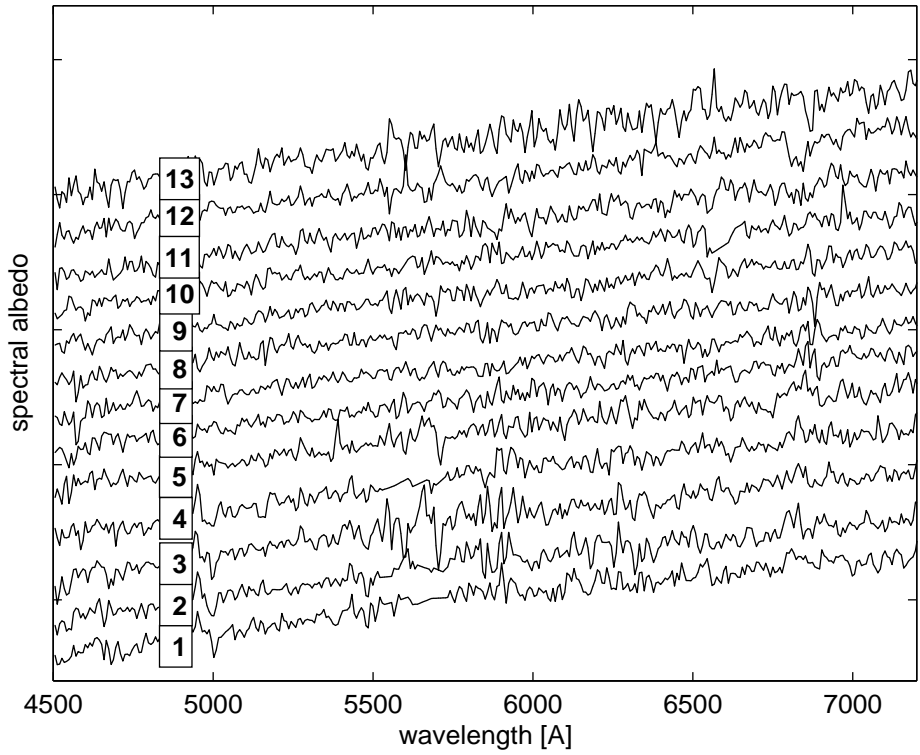

Fig. 10.

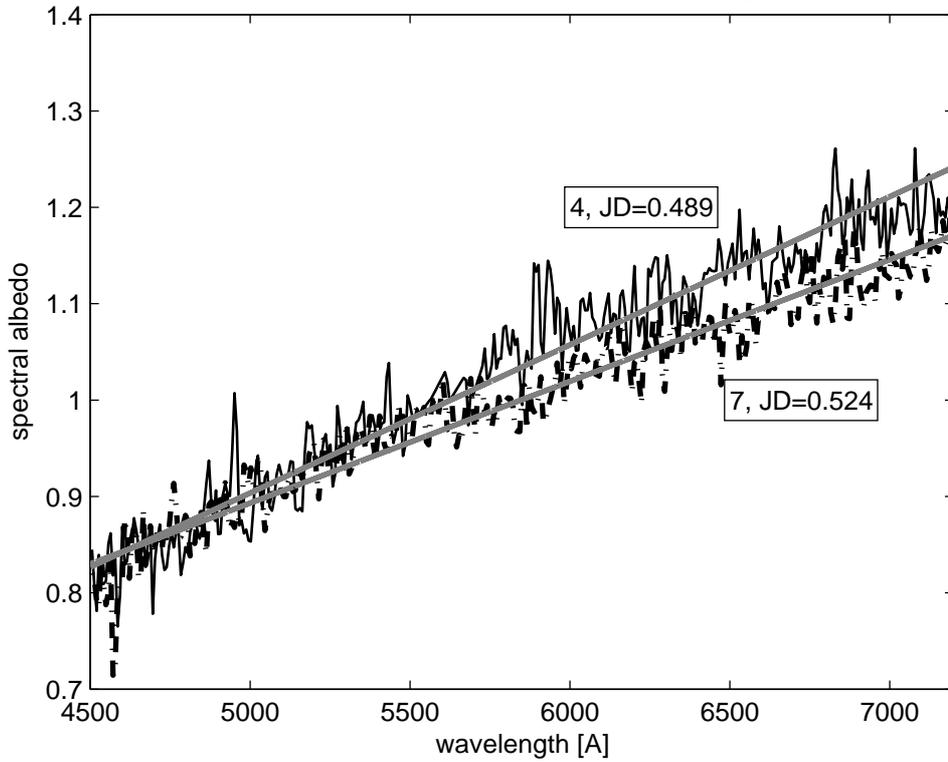

Fig. 11.

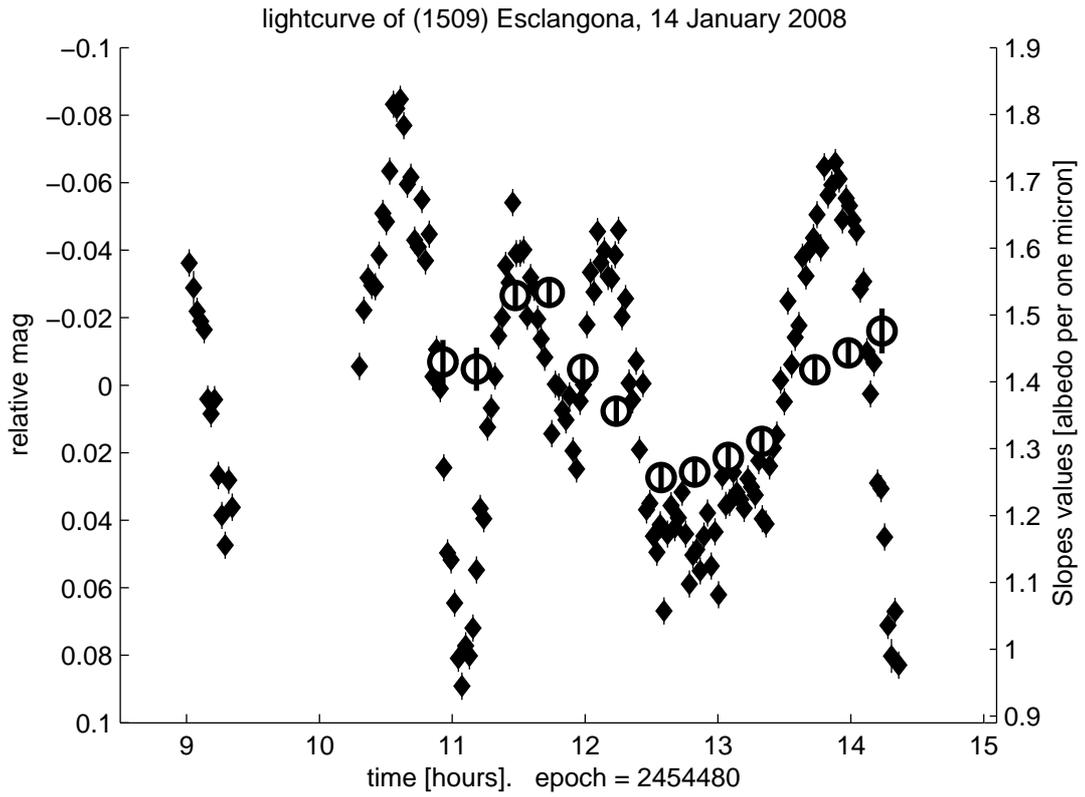

Fig. 12.

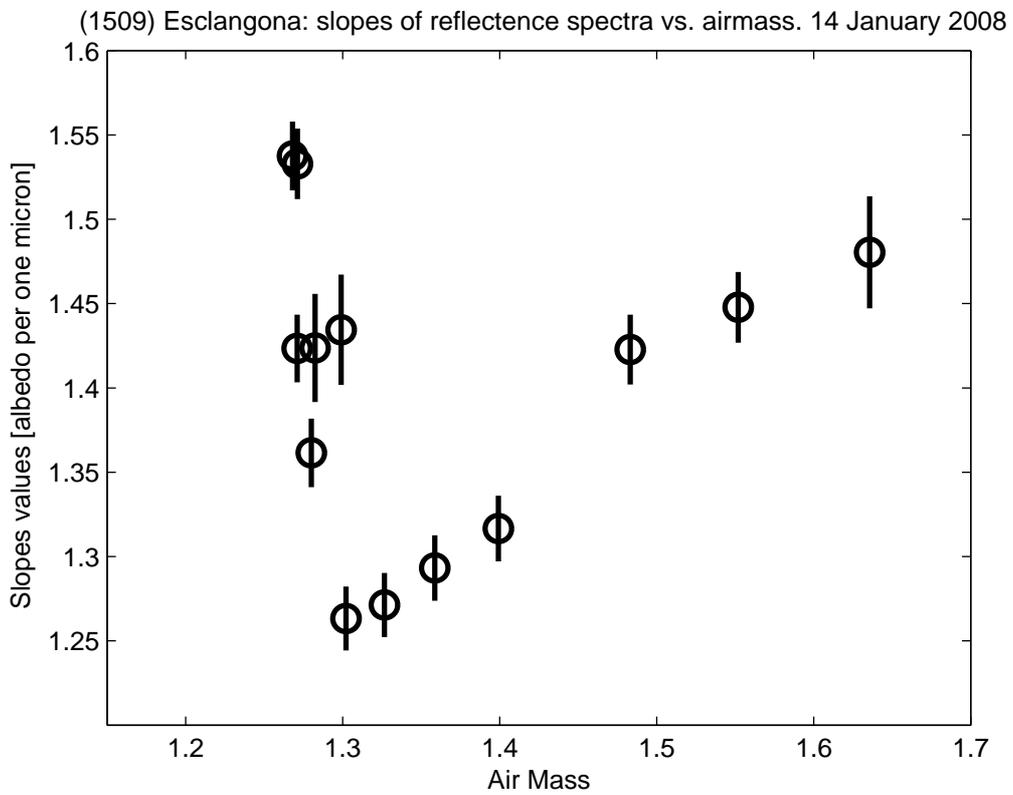

Fig. 13.

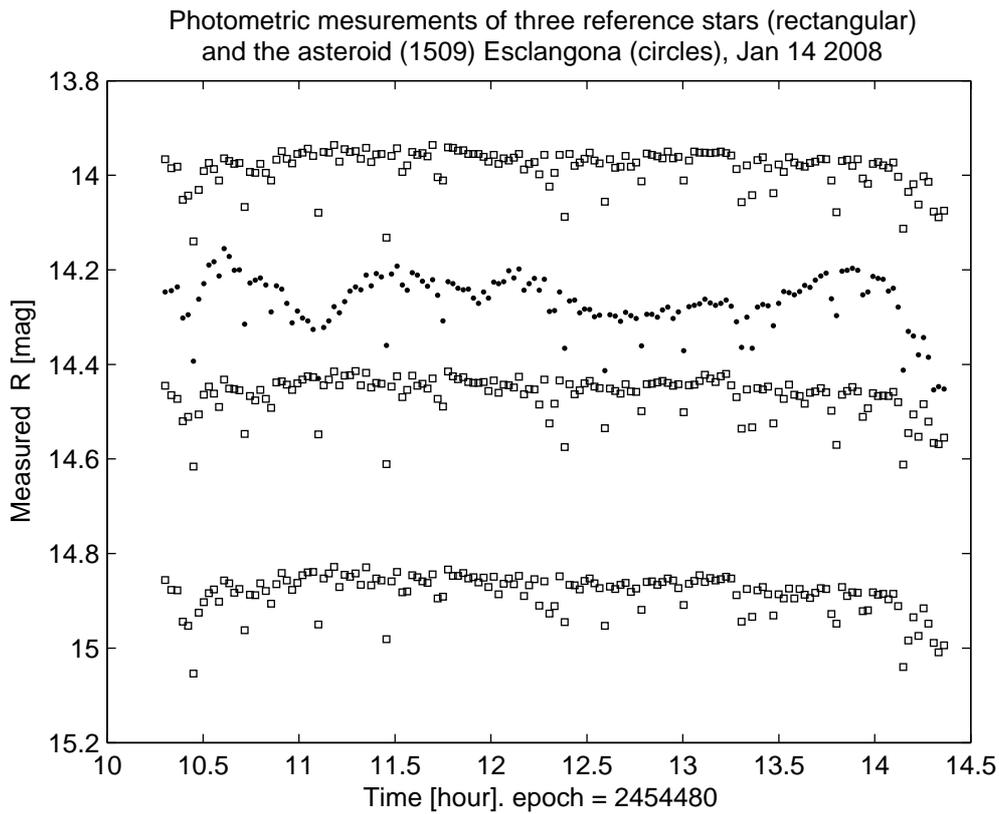

Fig. 14.

**Figures Captions:**

Fig. 1: spectra of *(90) Antiope* and *(1509) Esclangona* compared to data from the literature: The Small Main-belt Asteroid Spectroscopic Survey (SMASS) covers a wavelength range of 0.435-0.925 μm (Bus and Binzel 2002); The Eight-Color Asteroid Survey (ECAS) was done with eight narrow band filters (Zellner et al. 1985).

Fig. 2: *(90) Antiope* photometric lightcurve (black points) and the corresponding spectra fluxes (white rectangular) drawn on it, January 11$^{th}$, 2008. The running numbers match the markings on Fig. 3. The three gray points on top of the lightcurve were taken from Fig. 10 of Descamps et al. (2007) and represent the beginning, middle and end of the eclipse. As displayed, the match of our measurements to the lightcurve of Descamps et al. (2007) is full, meaning that our observations took place while the eclipse was almost total, as can be learned from Descamps et al. (2007) model.

Fig. 3: The 12 spectra of *(90) Antiope* obtained on January 11th. The spectra were shifted in the Y-axis for display. The running numbers match the markings on Fig. 2.

Fig. 4: Some spectra examples of *(90) Antiope* from January 11th while the eclipse took place. The spectra are offset by the relative brightness of the signal during the mutual event, thus they are displayed in real ratio. The running numbers match the markings on Fig. 2.

Fig. 5: The 12 spectra of *(90) Antiope* obtained on January 12th. The spectra were shifted in the Y-axis for display. The running numbers match the markings on Fig. 6.

Fig. 6: *(90) Antiope* photometric lightcurve (black points) and the corresponding spectra fluxes (white rectangular) drawn on it, January 12th, 2008, while the secondary eclipse took place. The running numbers match the markings on Fig. 5.

Fig. 7: Comparison between two spectra of *(90) Antiope* from the two nights, showing a lack of differences between the two components' spectra.

Fig. 8: The lightcurve of *(1509) Esclangona* from three nights folded with a period of 3.2524 hours. The unusual features on the lightcurve repeat on different nights.

Fig. 9: *(1509) Esclangona* photometric lightcurve (black points) and the corresponding spectra fluxes (white rectangular) drawn on it, January 14th, 2008. The running numbers match the markings on Fig. 10.

Fig. 10: The 13 spectra of *(1509) Esclangona* obtained on January 14th. The spectra were shifted in the Y-axis for display. The running numbers match the markings on Fig. 9.

Fig. 11: Reflectance spectra (black line and dashed line) of *(1509) Esclangona* with different slopes (grey lines) displaying alteration of the asteroid's color. The two reflectance spectra were drawn on each other in the blue side to clear the slope difference in the red side. The running numbers match the markings on Fig. 9.

Fig. 12: Lightcurve of *(1509) Esclangona* from January 14th (black points), with the corresponding slopes of the reflectance spectra (white circles). The "slope curve" was scaled to fit the Y axis for display. The change of the slope is continuous and fits a specific location on the lightcurve – the wide and flat minimum, around 13 hours.

Fig. 13: The slopes of the reflectance spectra (in units of albedo per micron) against the air mass of each spectrum of *(1509) Esclangona*. The non-linearity of the plot shows that the slopes are not correlated with the telescope's elevation.

Fig. 14: The photometric lightcurves of three reference stars (empty rectangular) located in the same field of view of the *C18* as *(1509) Esclangona* (black points). These lightcurves were not calibrated hence they reflect the atmospheric stability. The brightness of the nearby stars remained steady while the slope of the reflectance spectra changed (especially between 12 to 14 hours, compared to Fig. 12).